\documentclass[12pt,oneside, a4paper]{revtex4}
\pdfoutput=1
\usepackage{graphicx}
\usepackage{amssymb}
\usepackage{amsmath,amssymb}
\usepackage{color}
\usepackage{setspace}
\usepackage{fancyhdr}
\usepackage{multirow,natbib}
\usepackage{caption}
\usepackage{subcaption}

\usepackage{xcolor}
\usepackage{todonotes}

\definecolor{darkblue}{rgb}{0,0,0.5}
\usepackage[colorlinks,linkcolor=blue,citecolor=blue,urlcolor=blue]{hyperref}

\allowdisplaybreaks

\usepackage[utf8]{inputenc}
\usepackage{newunicodechar}


\newcommand\beq{\begin{equation}}
\newcommand\eeq{\end{equation}}

\usepackage{xcolor}
\usepackage{todonotes}

\begin{document}
\title{Fragmentation Functions:\\ Modifications and new Applications}
\date{\today}
\author{A.K.~Likhoded}
\email{Anatolii.Likhoded@ihep.ru}
\affiliation{A. A. Logunov Institute for High Energy Physics, NRC KI,
Protvino, RF}
\author{V.A.~Petrov}
\email{Vladimir.Petrov@ihep.ru}
\affiliation{A. A. Logunov Institute for High Energy Physics, NRC KI,
Protvino, RF}
\author{V.D.~Samoylenko}
\email{Vladimir.Samoylenko@ihep.ru}
\affiliation{A. A. Logunov Institute for High Energy Physics, NRC KI,
Protvino, RF}
%
%----------------------------------------------------
\begin{abstract}
We present a heavy quark fragmentation mechanism based on a modified version of the standard model of quark fragmentation, taking into account peculiarities of the Regge trajectories of quarkonia containing heavy quarks. The modified model describes the experimental data more satisfactorily both for meson and baryon production.
\end{abstract}
%-------------------------------------------------------
%
\maketitle
\section*{Introduction}
Production of heavy hadrons ($H$) in $e^{+}e^{-}$ annihilation is usually viewed as a convolution
of a perturbative part (hard gluon emission at energies above $\approx 1$~GeV)
and a nonperturbative part, called hadronization or fragmentation, in which quarks
are confined into colorless hadrons.
Given this definition, the fragmentation process can be described in terms of
the probability of producing a hadron $H$ with a given value $z$, called $D_{q}^{H}(z)$ (or $D(x_H)$),
where $q$ is the flavour of the generating quark.
Measuring $D(x_{H})$ at different energies can be used to test the evolution of QCD,
and comparing $D(x_B)$ and $D(x_D)$ can serve as a test of the symmetry of heavy quarks.
One of the first theoretical predictions about the difference in inclusive spectra for light and heavy quarks was made in
\cite{Kartvelishvili:1977pi}.
Experimental confirmation of the differences in the distribution of inclusive spectra for light and heavy quarks was presented in \cite{Chliapnikov1980},
which analyzed the inclusive production of vector mesons in $K^{+}p$ interactions at 32 GeV/c and obtained inclusive cross sections for these processes.
It was noted that the behavior of the strange quark $s$ differed from that of the light quarks $u$, $d$.

In \cite{Chliapnikov:1977fc}, the authors used different Regge trajectories for different quark flavours, which made it possible to describe the distribution of valence quarks in $\pi$-, $K$-, and $D$-mesons. Furthermore, based on the quark fusion mechanism, inclusive spectra of vector mesons produced in $K^+ p$ interactions at 32~GeV/c were calculated, and their good agreement with experimental data was shown.

In the same work, experimental values obtained for the cross section for inclusive production of vector mesons, with a clear dominance of the $\phi$-meson production compared to the $\rho$-meson production, which is in direct agreement with the violation of
$SU(3)$ symmetry in the $Kp$ interaction that we are considering.
In this paper, we present a fragmentation model based on a modification of the model
\cite{Kuti:1971ph} and 
which provides more insight into the quark-gluon structure of the hadrons under study.
The mentioned modification in the presence of heavy quarks \cite{Seuster_2006} consists of replacing the light quark parameters
($\alpha_v$ and $\gamma_M$) with the parameters of heavy quarks in accordance with their Regge trajectories. This approach allows us to describe the fragmentation of a heavy quark, of course
within some assumptions present in the original model.
Our model allows us to describe the structure functions of hadrons within the Bjorken formalism,
within which the distribution functions of valence quarks and gluons
in the hadron under study are naturally derived.
\section*{Fragmentation Function}
We will conduct our reasoning within the framework of the parton model, where the structure function $W(x)$ for
small values $x = \frac{Q^2}{2M \nu}$ looks as follows
\begin{align}
W(x) = A + \sum_i B_i x^{-\alpha_i(0)},
\label{eq:stfun}
\end{align}
where $A$ is the Pomeron exchange parameter,
$\alpha_i(0)$ is the Regge trajectory intercept for the parton, and $B_i$ is the normalization coefficient. For convenience, we will henceforth use the abbreviated notation $\alpha_i \equiv \alpha_i(0)$.
The sum $\sum_i B_i x^{-\alpha_i(0)}$ in (\ref{eq:stfun}) is related to the contribution of valence quarks.
The value of the Regge trajectory intercept $\alpha_i$ associated with quark $q_i$ depends on the quark type: for heavier quarks, the intercept is located lower than for lighter quarks.
Thus, within the parton model, for small values of $x$ the behavior of the structure function is determined by the intercept parameter $\alpha_i$,  where $\alpha_i(0)$ is the Regge trajectory intercept for the parton.

Estimates of the parameter $\alpha_i$ for different quarks $i$ show that they vary within the following limits:
\begin{align}
    \begin{split}
        \alpha_u &\approx 0.5, \\
        \alpha_s &= 0:0.2, \\
        \alpha_c &= -2.2:-3,  \\
        \alpha_b &= -9:-11
    \end{split}
\end{align}
The values of  $\alpha_c$ and $\alpha_b$ are poorly determined from Regge trajectories, which are not completely constructed for heavy quarks.
Below, we discuss the possibility of refining the values of these parameters.
In our  modified model, the meson structure function has the form
\begin{align}
    F_{v_1}^{M}(x) = N_M \  x^{-\alpha_{v_1}}  (1-x)^{-1 + \gamma_M + 1 - \alpha_{v_2}},
    \label{eq:kuti_M}
\end{align}
while for baryon we get
\begin{align}
    F_{v_1}^{B}(x) = N_B \  x^{-\alpha_{v_1}}  (1-x)^{-1 + \gamma_B + 1 - \alpha_{v_2} + 1 - \alpha_{v_3}}.
    \label{eq:kuti_B}
\end{align}
Here $\alpha_{v_i}$ is the Regge trajectory intercept, and the index $v_i$ corresponds to the valence quarks. The constants $N_M$ and $N_B$ are chosen based on the normalization condition: the integrals $\int_0^1 F_{v_1}^M(x) dx$ and $\int_0^1 F_{v_1}^B(x) dx$ are equal to the number of corresponding quarks $v_1$ in the particle.
The values of the parameters $\gamma_M $ and $\gamma_B $ are determined by the behavior of the functions (\ref{eq:kuti_M}) and (\ref{eq:kuti_B}) in the region of large values of the variable $x \sim 1$ and are determined by the gluons accompanying the leading quark in mesons and baryons.
\section*{The relationship between the wave function of a hadron and its quark distribution function}
One of the key aspects of studying the structure of hadrons containing heavy quarks is the relationship between the particle's wave function and the quark distribution function. The latter was previously presented in formulas (\ref{eq:kuti_M}) and (\ref{eq:kuti_B}) and allows one to describe the properties of the quarks belonging to a given particle.

The relationship between the wave function of $\psi(0)$ particles containing heavy quarks and the quark distribution function was discussed in the paper \cite{Kartvelishvili:1985ac}.
Within the simplest model describing heavy quarks in the $Q\bar{Q}$ meson, the structure function in the valence quark approximation has the following form:
\begin{align}
    F_{Q}^{M}(x) \sim x^{-\alpha_Q} (1-x)^{- \alpha_{\bar{Q}}},
\end{align}
where $\alpha_Q$ ($\alpha_{\bar{Q}}$)  is the Regge intercept for $Q$ ($\bar{Q}$).
Note that $\alpha_Q$ = $\alpha_{\bar{Q}}$.
The relationship between the valence quark distribution function for quarkonium and the wave function is defined as
\begin{align}
   F^M(x)
  &  = \int d^3 q \, |\psi(\vec{q})|^2 \delta \left( x - \frac{q_0 + q_z}{M} \right)  \nonumber \\
  &  = \frac{\pi M^3}{4}\int\limits_{\phi(x)}^{1} dv^2 |\psi(v^2)|^2
\end{align}
\begin{align}
|\psi(v^2)|^2
&= -\frac{4}{\pi M^3} \frac{\mathrm{d}f\left(\phi^{-1}(v^2)\right)}{\mathrm{d}v^2} \nonumber \\
&= \frac{-\alpha_Q}{\pi M^3 B\left(1 - \alpha_Q,\,1 - \alpha_Q\right)} \left(\frac{1 - v^2}{4}\right)^{-\alpha_Q - 1}
\label{eq:kart8}
\end{align}

Equation (\ref{eq:kart8}) allows one to determine the parameter $\alpha_Q$ using the experimental width of the lepton decay
\begin{align}
    \Gamma \left( (Q\bar{Q}) \to e^+ e^- \right) &= 16 \pi \alpha_{em}^2 e_Q^2 \frac{|\psi(0)|^2}{M^2}
\end{align}
\begin{align}
    |\psi(0)|^2 &= \left| \frac{1}{(2\pi)^{3/2}} \int \psi(\mathbf{v}^2) \, d^3q \right|^2 = \frac{M^3 \Gamma(3/2 - \alpha)}{64 \pi \sqrt{\pi} \Gamma(-\alpha)}
    \left[ \frac{\Gamma\left( \frac{1-\alpha}{2} \right)}{\Gamma\left( \frac{4-\alpha}{2} \right)} \right]^2
\end{align}
By comparing the lepton widths in the $J/\psi$ and $\Upsilon$ cases, we can calculate the values of $\alpha_c $ and $\alpha_b$. Using modern values of the masses and widths \cite{ParticleDataGroup:2024cfk}, we obtain:
\begin{align}
    \begin{split}
        \alpha_c &= -2.93 \pm 0.07 \approx -2.9 \\
        \alpha_b &= -10.52 \pm 0.21 \approx -10.5
    \end{split}
\end{align}
\section*{Comparison of wave functions}
Next, we present further indirect confirmation of the relationship between the wave and structure functions of a particle.

The calculation of wave functions for heavy particles within the potential model is the subject of the work \cite{Bondar:2004sv}, which examined expressions for the wave functions of $J/\psi$- and $\Upsilon$-mesons in the valence quark approximation. The authors' primary motivation was the discrepancy between the experimental results of Belle \cite{Belle:2002tfa} and the NRQCD predictions \cite{Braaten:2002fi} for
the cross section of the process $e^+ e^- \rightarrow J/\psi + \eta_c$,
caused primarily by the use in NRQCD of a simplified function $\delta(x - 0.5)$ instead of
the charmonium wave function,
which significantly lowered the predictions compared to experiment.
In the work \cite{Bondar:2004sv} the wave function is described as follows
\begin{align}
    \phi_o(x, v^2) = c_o(v^2) x_1 x_2 \left\{ \frac{x_1 x_2}{\left[1 - 4 x_1 x_2 (1 - v^2) \right]} \right\}^{1 - v^2},
\end{align}
\begin{align}
    \int_{0}^{1} dx_1 \, \phi_o(x, v^2) = 1.
\end{align}
where $x_1, x_2$ are the fractions of the meson momentum carried by the quarks, $v$ is a parameter representing the characteristic velocity of a quark in a bound state, and $c_0(v^2)$ is a normalization constant.

As stated earlier, in the valence quark approximation within our model, the quark distribution function takes the following form:
\begin{align}
    F_{q_i}^{M}(x) = N_M \  x^{-\alpha_v} (1-x)^{- \alpha_s}
    \label{eq:kuti_M_valentn}
\end{align}
Here we take into account that in the valence quark approximation, $\gamma_M = 0$ and this allows us to construct distribution functions functions unambiguously.
Let us compare the predictions of our model in the valence quark approximation with the results obtained in \cite{Bondar:2004sv}.
\begin{figure}[h]
    \centering
    \begin{subfigure}{0.45\textwidth}
        \centering

        \includegraphics[width=\linewidth]{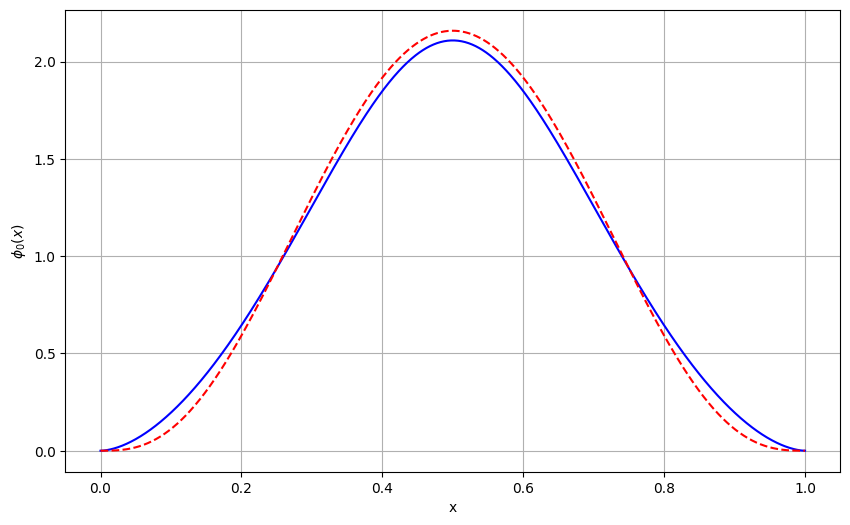}
        \caption{Charmonium}
        \label{fig:cc_bondar_chernyak}
    \end{subfigure}
    \hfill
    \begin{subfigure}{0.45\textwidth}
        \centering
        \includegraphics[width=\linewidth]{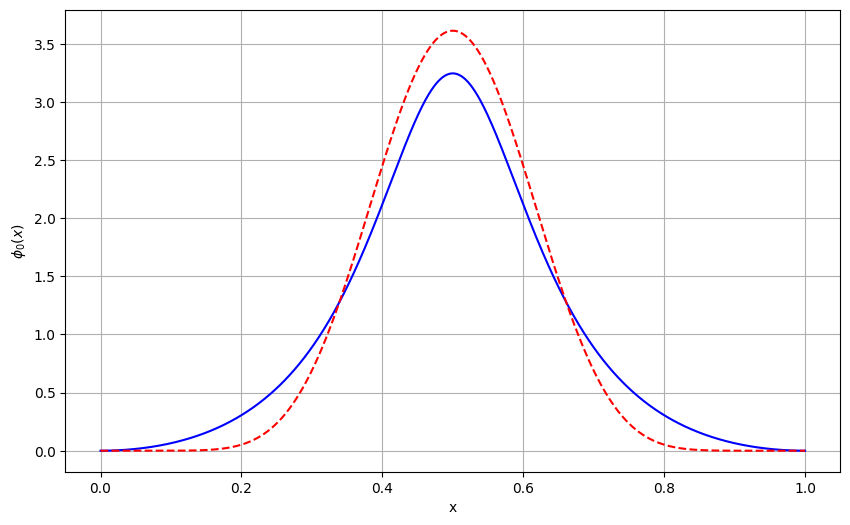}
        \caption{Bottomonium}
        \label{fig:bb_bondar_chernyak_fixed}
    \end{subfigure}
    \caption{Comparison of particle wave functions in our approach (red lines) and in the model \cite{Bondar:2004sv}  (blue lines)}
    \label{fig:qq_bondar_chernyak}
\end{figure}
As can be seen from Fig.~\ref{fig:qq_bondar_chernyak}, the two different theoretical approaches demonstrate agreement. Remarkably, our model, using only one parameter, was able to successfully describe the particle's wave function, whereas in \cite{Bondar:2004sv} this requires a significantly more complex model. This underscores the effectiveness of our proposed approach and its ability to reproduce key hadron characteristics with a minimal number of free parameters.

We previously showed that the quark function in a hadron is related to its wave function.
Thus, the wave function influences the distribution of the particle production cross section by the fraction of momentum transferred, $x$. Note that this distribution also depends on the quark source.

The fragmentation function allows us to describe the influence of the hadron source:
\begin{align}
    f(x) = \int_x^1 \phi(y) F(\frac{x}{y}) dy
\end{align}

Here, $y$ is the fraction of the momentum of the fragmenting quark; $x$ is the fraction of the momentum carried away by the final particle; $\phi(y)$ is the distribution of fragmenting quarks in the initial state, $F(\frac{x}{y})$ is the squared wave function of the final particle.
The use of such a fragmentation function made it possible to successfully describe the experimental data in the paper \cite{Gerasimov:2023ttp}.

In the case of production in $e^+e^-$ annihilation, the distribution of initial quarks is described by
the $\delta$-function, so after integration, the fragmentation function exactly coincides with the
structure function. Thus, in this case, formulas (\ref{eq:kuti_M}) and (\ref{eq:kuti_B}) can be used to
calculate the fragmentation function.

\section*{Production of mesons with a heavy quark}
\begin{figure}[h]
\begin{minipage}{0.47\linewidth}
\center {\includegraphics[width=\linewidth]{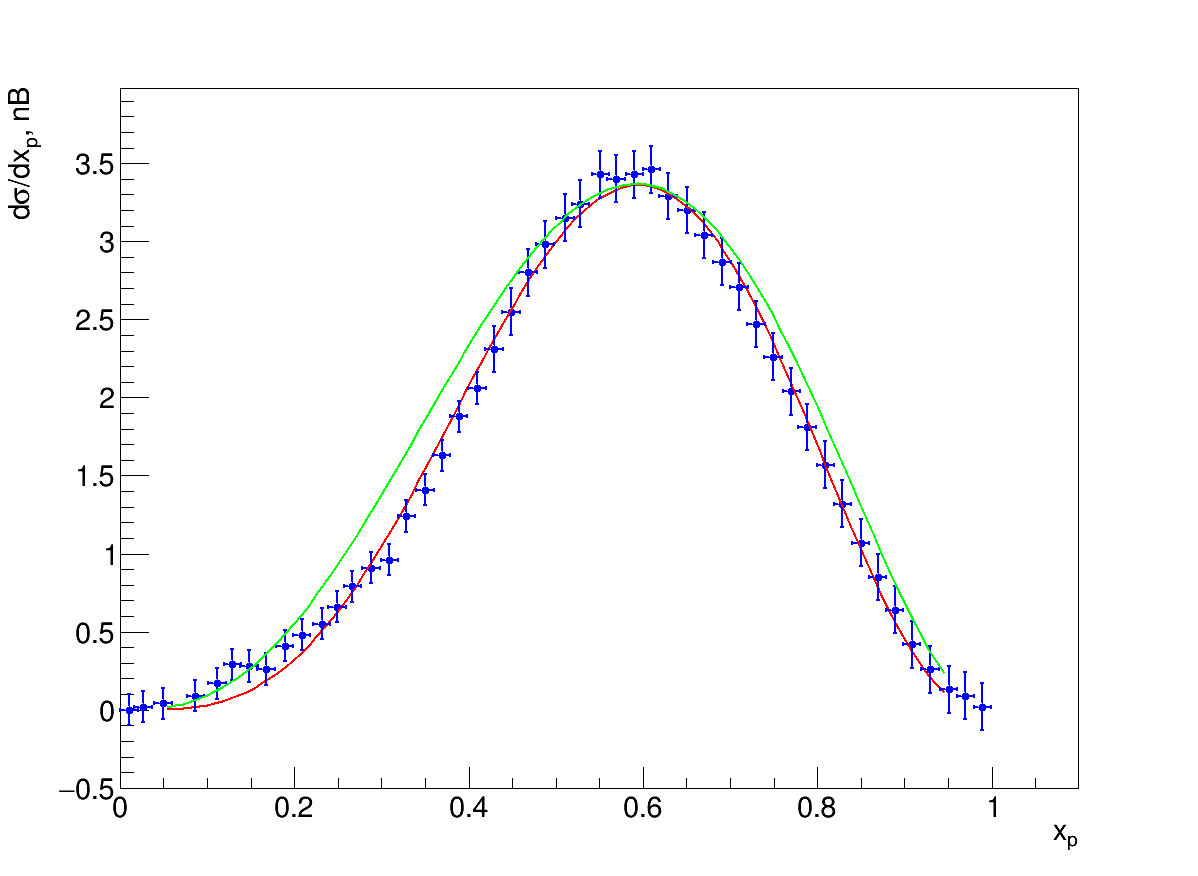}}
\end{minipage}
\begin{minipage}{0.47\linewidth}
%\center {\includegraphics[width=\linewidth]{pics/Fig_D0_CLEO_fit.png}}
\center {\includegraphics[width=\linewidth]{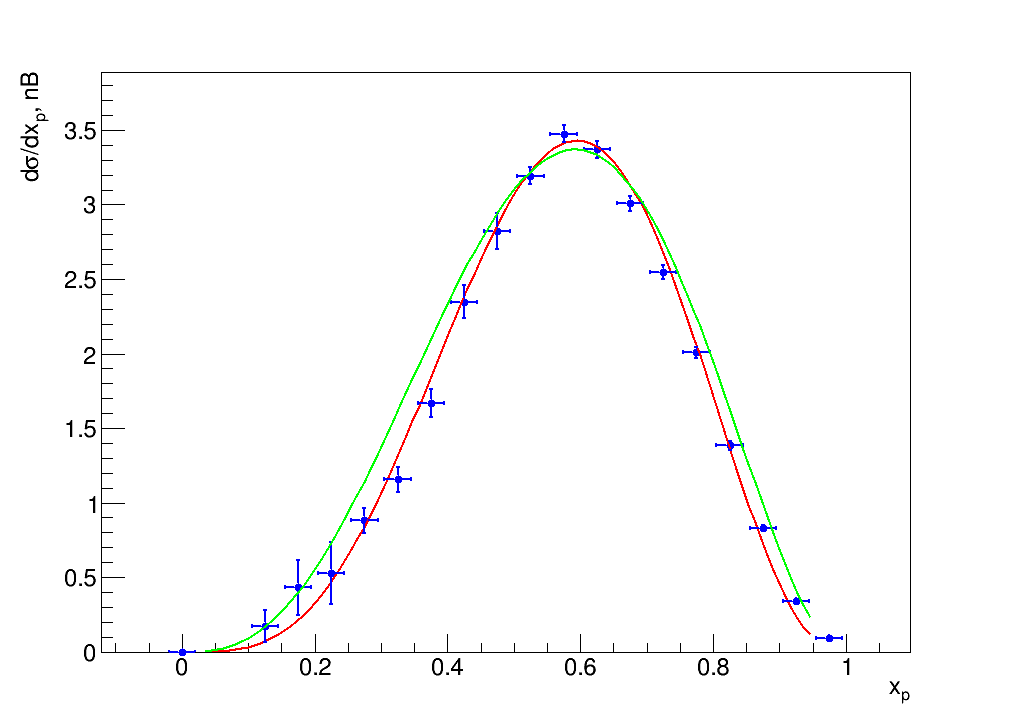}}
\end{minipage}
\caption{Fit by fragmentation function (\ref{Eq_FF_D0})  $D^0$-meson production (Data  of Belle is shown on the left and CLEO
on the right). 
Green line is the fragmentation function.}
\label{fig_D0}
\end{figure}

Consider the production of a $D0$ meson in $e^{+}e^{-}$ annihilation. This process has been studied in many experiments; we will consider the results of Belle \cite{Seuster_2006} and CLEO \cite{Artuso_2004}.
As noted above, in the case of $e^{+}e^{-}$ annihilation, the fragmentation function exactly coincides
with the wave function of the final particle.
Fig.~\ref{fig_D0} shows a comparison of the theoretical predictions of our model with the results of Belle
and CLEO, which are very close.
The following parameters are used for the theoretical prediction:
\begin{align*}
    \begin{split}
        \alpha_c &= -2.9 \\
        \gamma_M &= 2.5
    \end{split}
\end{align*}
Then the fragmentation function takes the form
\begin{align}
    f(x) \sim x^{2.9} (1-x)^{2}.
\label{Eq_FF_D0}
\end{align}
%
%-----------------------------------------------
%
\begin{figure}[htb]
\center {\includegraphics[width=80mm]{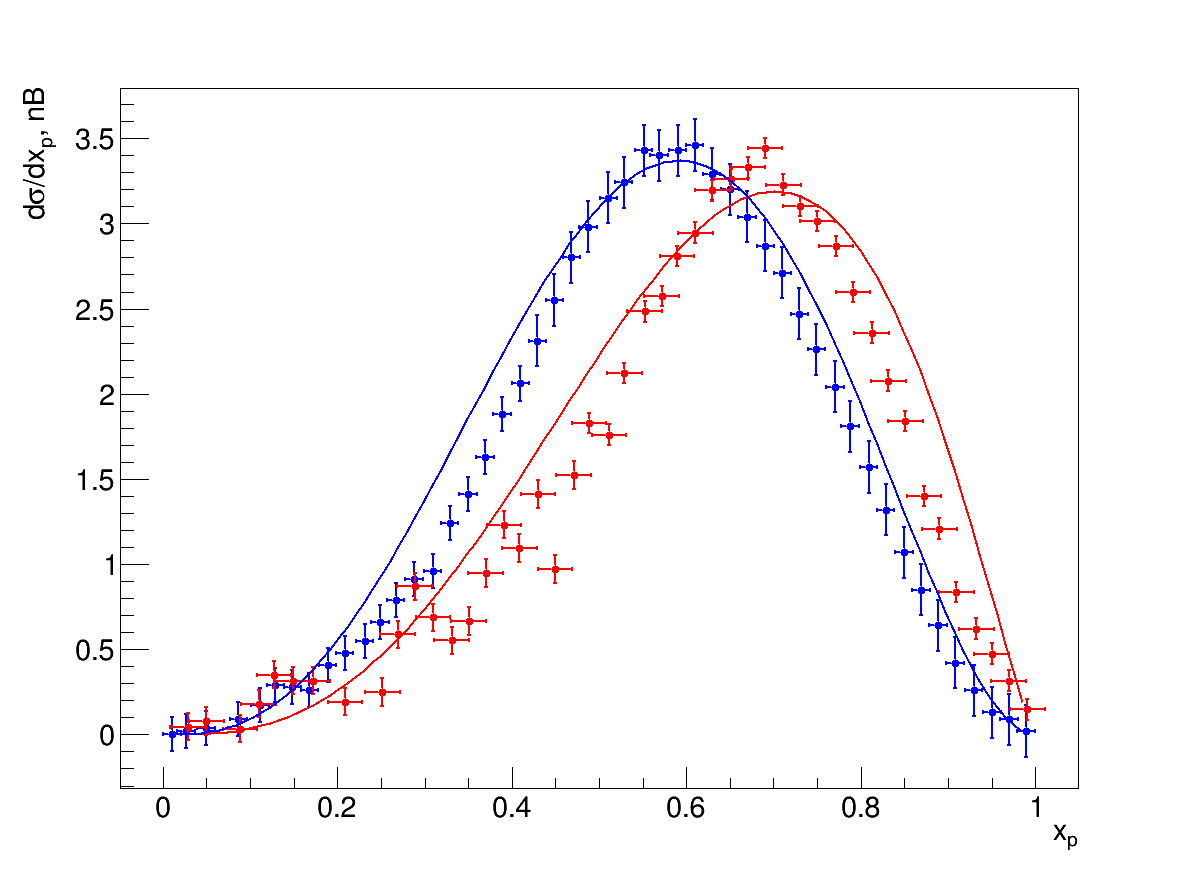}}
\caption{Description of the fragmentation of the $D_s$-meson production function
(\ref{eq:kuti_M_valentn}) based on Belle experiment.
Blue color for the $D^0$-meson, red for the $D_s$-meson. For clarity, the $D_s$-meson cross section
is enlarged.
}
\label{fig_Ds}
\end{figure}
%
%--------------------------------------------------------------------
%

As can be seen from Fig.~\ref{fig_D0}, our theoretical predictions demonstrate good agreement
with the experimental data.
It should be noted that the parameter $\gamma_M = 2.5$ has a significant uncertainty.
Fig.~\ref{fig_Ds} shows the fragmentation of the $D_s$ meson with an exponent of $(1 - x)^{1.3}$, which
is $\approx 0.5$ smaller than in formula (\ref{Eq_FF_D0}) and agrees with the theoretical prediction.
For comparison, the already discussed fragmentation of the $D^0$ meson is shown.
From this figure, one can see the difference between the wave functions of the $D^0$ and $D_s$ mesons, i.e. the difference when replacing a light antiquark with a strange antiquark
(recall that the quark structure of the $D^0$ meson is $c \bar {u}$, and
that of the $D_s$ is $c \bar {s}$).

\begin{figure}[htb]
\begin{minipage}{0.47\linewidth}
\center {\includegraphics[width=\linewidth]{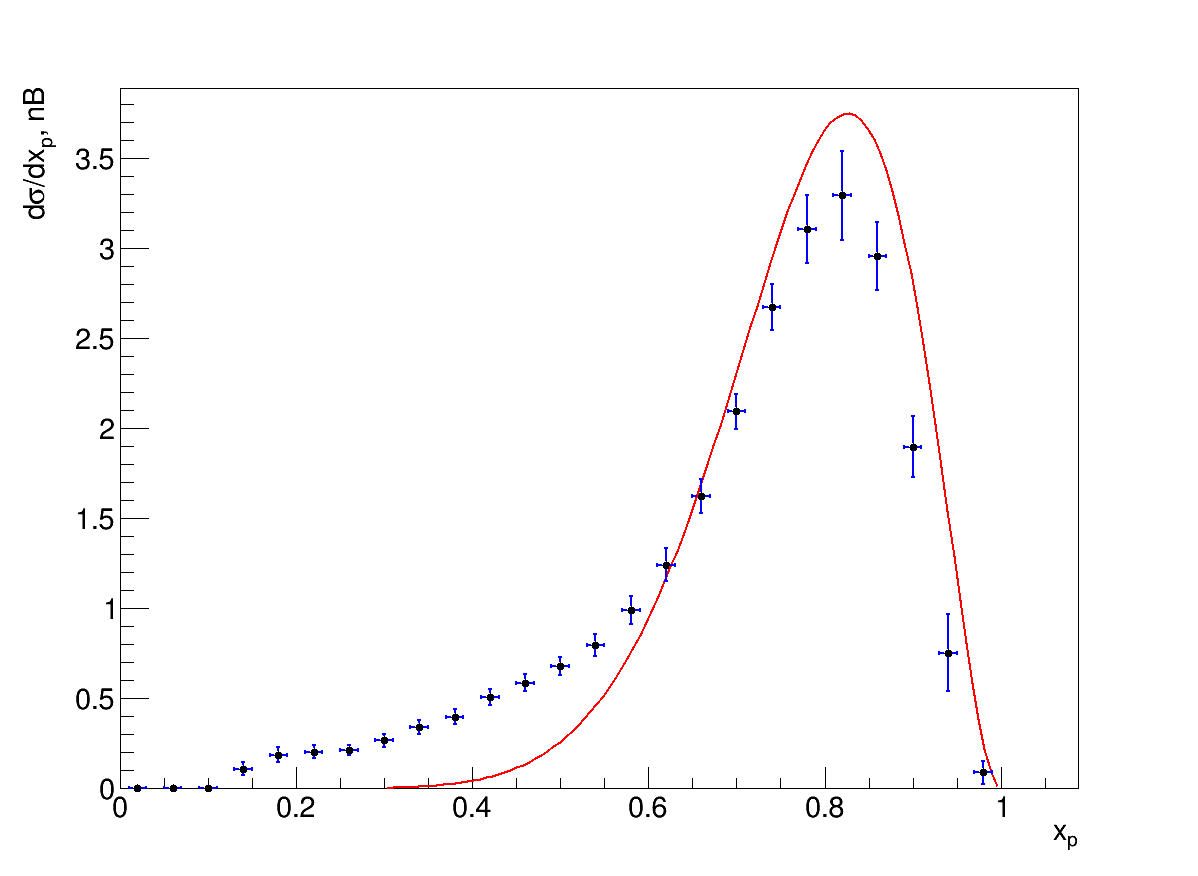}}
\end{minipage}
\begin{minipage}{0.47\linewidth}
\center {\includegraphics[width=\linewidth]{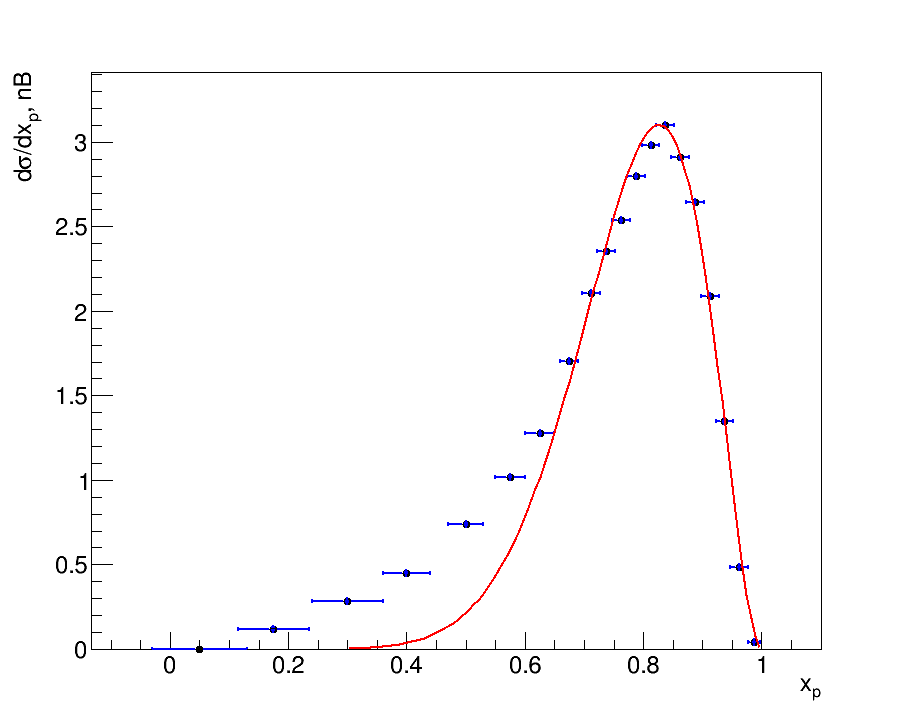}}
\end{minipage}
\caption{Description of the $B$-meson production distribution by the fragmentation function (\ref{Eq_FF_B}) based on data from the SLD (left) and ALEPH (right) experiments.}
\label{fig_B}
\end{figure}

Now let's consider the fragmentation of the heavier $b$ quark. We will rely on the results of the
ALEPH \cite{ALEPH:2001pfo} and SLD \cite{SLD:2002poq} experiments, which studied the production
of $B$ mesons in $e^+e^-$ annihilations.
Fig.~\ref{fig_B} shows a comparison of the theoretical predictions within our model with the experimental results.
The following parameters are used for the theoretical prediction:
\begin{align*}
    \begin{split}
        \alpha_b &= -9.5 \\
        \gamma_M &= 2.5
    \end{split}
\end{align*}
In this case, the fragmentation function takes the form
\begin{equation}
f(x) \sim x^{9.5} (1-x)^{2}
\label{Eq_FF_B}
\end{equation}
%-----------------------------------------------------
As with the $D$-meson, good agreement with experiment is observed. Note the discrepancy between our model and the experimental data in the region of small x values.
\section*{Baryon Production}
\begin{figure}[htb]
    \centering
   \includegraphics[width=80mm]{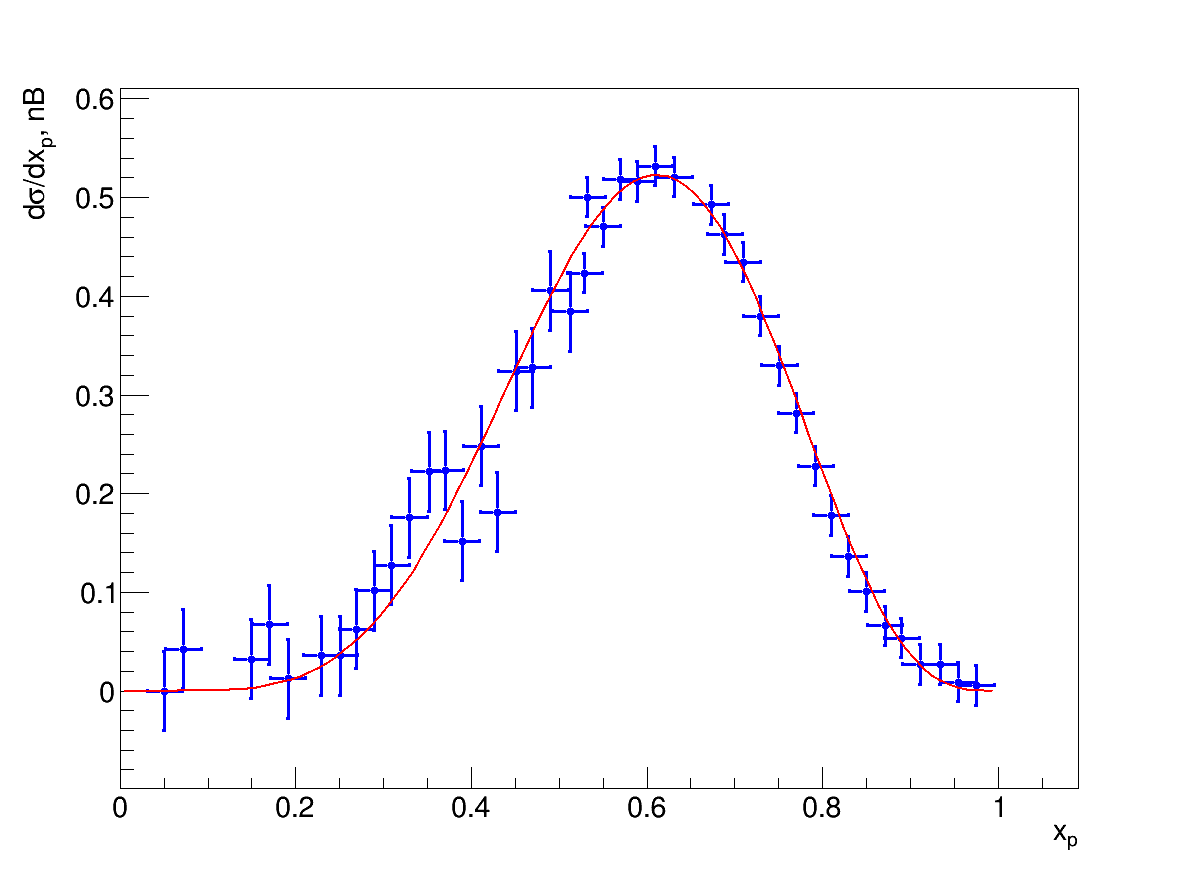}
\caption{Comparison of fragmentation functions for the $\Lambda_c$-baryon with experimental Belle data.}
\label{fig:lambda_c}
\end{figure}

Let's consider baryon production, in particular the production of the $\Lambda_c$ baryon as a result of $e^+e^-$ annihilation. Fig.~\ref{fig:lambda_c} compares the theoretical predictions of our model (blue curve) with the results of the BELLE experiment \cite{Seuster_2006} (red crosses).

The following parameters are used for theoretical prediction:
\begin{align*}
    \begin{split}
        \alpha_c &= -2.9 \\
        \gamma_B &= 2.5
    \end{split}
\end{align*}
Then the fragmentation function takes the form
\begin{align*}
    f(x) \sim x^{2.9} (1-x)^{2}
\end{align*}
%
%---------------------------------------------
\section*{Conclusion}
The nonperturbative part of the quark fragmentation function, which we discuss in our paper, satisfactorily
describes experimental data on hadron production with heavy quarks in $e^+e^-$ annihilation.
Moreover, the functional dependence of the spectrum of the produced hadron in this reaction is determined by its quark distribution function,
which, in turn, is determined by our modified model with a different treatment of valence quarks (Taking this opportunity, we draw attention to a recent work \cite{Kiselev:2025lpb}  that provides an original interpretation of the mechanism of fragmentation of light-heavy quarkonia). The latter is determined by the change in the parameter responsible for the intercept  of the Regge trajectory
associated with the valence quark ($\alpha$) in the formula (\ref{eq:stfun}) for the structure function $W(x)$ at small $x$.

The model we are considering is quite simple and somewhat reminiscent of the equivalent photon method in quantum electrodynamics. It takes into account only the contributions of valence quarks and gluons. Varying these contributions
depending on the Regge parameters $\alpha$ and the gluon contribution allows us to reproduce the specific features of the fragmentation functions of heavy
quarks and obtain a satisfactory description. 

Another conclusion we draw is that the specific features of hadrons
under the condition of confinement of their constituent quarks suggests the presence of a wave function in the produced hadrons, a particular example of which, in the case of heavy quarkonium, we present in our paper.

\bibliographystyle{unsrt}  % Choose a style: plain, abbrv, unsrt, etc.
%\bibliography{references}
\bibliography{references}

\end{document}